# Phase-matched electron-photon interactions enabled by 3D-printed helical waveguides

Masoud Taleb,[1] Mohsen Samadi,[2] and Nahid Talebi[1,*]

[1]Institute of Experimental and Applied Physics, Kiel University, 24098 Kiel, Germany
[2]Department of Electrical and Information Engineering, Kiel University, 24143 Kiel, Germany

**ABSTRACT**. The Smith-Purcell effect enables electromagnetic radiation across arbitrary spectral ranges by phase-matching the diffraction orders of an optical grating with the near-field of a moving electron. In this work, we introduce a novel approach using a helically shaped waveguide, where phase-matching is achieved through guided light within a helical optical fiber fabricated via two-photon polymerization using a 3D printer. Our results demonstrate that radiation from these structures precisely satisfies the phase-matching condition and is emitted directionally at specific angles, contrasting with the broad angular distribution characteristic of the traditional Smith-Purcell effect. Helical electron-driven photon sources establish a new paradigm, enabling 3D-printed structures to control electron-beam-induced radiation and, inversely, to facilitate light-induced efficient electron beam shaping and acceleration.

## Introduction

Electron-driven photon sources are electromagnetic radiation sources integrated into electron microscopes as internal emitters [1-4], with applications in spectral interferometry [5, 6] and time-resolved spectroscopy [7]. The main advantage of these sources is that they allow for generation of optical fields that are mutually coherent with respect to the evanescent near-field of the moving free-electron in vacuum and can be used for Ramsey-type interferometry of quantum objects in two-dimensional materials [8]. To realize such experiments, it is essential that the emitted light propagates collinearly with the electron trajectory, thereby enabling simultaneous excitation of the sample by both the generated light from the electron-driven photon source and the electron beam.

These sources typically consist of nanostructured metallic planar films that in interaction with electron beams generate light with specific radiation profiles, such as focused radiation [1] or vortex beams [9, 10]. While they provide a versatile platform for generating shaped electromagnetic radiation at specific photon energies defined by surface plasmon resonances, their relatively weak radiation intensities limit their effectiveness in nonlinear optics and pump-probe spectroscopy applications.

Hence, other radiation mechanisms from electron beams could be utilized to enhance the radiation intensity, particularly through phase-matching (synchronization) between the near-field of the moving electron and the induced polarizations in the material [11-16]. Cherenkov radiation [17] and Smith-Purcell effect [18] are among radiation mechanisms enabled by phase-matching.

Cherenkov radiation is the radiation caused by electron beams travelling in a medium where the phase velocity of light is larger than the electron group velocity. Cherenkov radiation forms a light cone along the electron trajectory, with the cone angle $\theta$ specified by the phase-matching condition $nk_0 \cos\theta = \omega v_e^{-1}$, resulting in $\cos\theta = c(nv_e)^{-1}$ [3, 19]. Here, $k_0$ is the radiation wavenumber, $\omega$ is its angular frequency, and $v_e$ is the electron's group velocity. This criterion however, neglects the fact that electron beams cannot propagate for long distances in materials, due to the strong elastic and inelastic scatterings from atoms and material excitations. Instead, electron beams travelling in vacuum near a vacuum/material interface and parallel to it provide an alternative [20, 21]. When the Cherenkov radiation is coupled to a waveguide mode copropagating with the electron, it leads to a strong, resonant, and phase-matched electron-photon interaction [22, 23]. However, the Cherenkov radiation in this case remains in the waveguide, making it inaccessible for far-field detectors.

Smith-Purcell effect mediates electron-photon interactions via coupling to optical grating modes, with a discrete set of wave numbers as $k_m = k_0 \cos\theta + 2m\pi\Lambda^{-1}$, associated with optical rays propagating in various directions in vacuum [24-26]. Here, $\Lambda$ is the grating period and $m$ is an integer, defining the diffraction mode. The phase-matching scenario in this case is recast as $k_m = \omega v_e^{-1}$. Generating coherent Smith–Purcell radiation with a tailored beam profile—such as a Gaussian beam aligned with the electron propagation direction—requires the use of cylindrical gratings or curved metamaterial structures, which present significant challenges for current nanofabrication technologies [27, 28].

Here, we demonstrate that phase-matching between a moving electron and the generated optical waves can be achieved using a waveguide shaped into a helical configuration. This geometry effectively extends the optical path length within the waveguide, thereby facilitating sustained phase-matching over longer interaction distances. Furthermore, the emitted light remains collinear with the electron trajectory and preserves a circular polarization, which is an important advantage for performing phase-locked photon–electron spectroscopy on two-dimensional materials, such as probing chiral excitons in transition metal dichalcogenides [29].

**Results and Discussions**

A 3D-printed helical waveguide is proposed and explored here (Fig. 1(a)), that allows for the phase-matched generation of coherent light in interaction with a moving electron. The waveguide is composed of a polymer resin with the radius of 400 nm, where a thin layer of metal with the thickness of approximately 40 nm is deposited on it. The helix is fabricated using two-photon polymerization [30-35] in a Nanoscribe Quantum X system with a 63x NA1.4 objective and an IP-Dip 2 resin with the refractive index of $n_{pol} = 1.55$. The radius of the helix, measured as the distance between the center of the waveguide to helix axis, is $r_h = 1.76\,\mu\text{m}$, and the helical pitch is $\Lambda = 5.85\,\mu\text{m}$.

For simplicity, we assume that the waveguiding modes remain unchanged when the straight fiber is transformed into a helix. The waveguide dispersion can be solved analytically (Fig. 1(b); see Supplementary Note 1 for the solutions [36]). The modes of waveguides with cylindrical symmetries are generally decomposed versus the azimuthal order $n$, ranging from $n = 0$ (no azimuthal dependence; shown by blue lines in Fig. 1(b)) to higher order modes with $n \geq 1$ (shown by black and purple lines). The latter modes are hybrid in nature, i.e., they are in the form of the superposition of TE$_z$ (transverse electric to z) and TM$_z$ (transverse magnetic to z) modes. The fundamental mode of the waveguide is also a hybrid mode with $n = 1$ (purple line in Fig. 1(b)), lying outside the light line in polymer. This mode has a plasmonic nature, with its field being tightly confined to the thin metallic region (Fig. 1(b), right).

A moving electron propagating parallel to the helix axis at a distance of $r_h + b + 20\,\text{nm}$, therefore not travelling through the matter, can excite the propagating modes of the waveguide. The sequential interaction of electrons with the waveguide at each turning point of the helix leads to the sequential emission of photons into guided modes. The optical path taken by the photons travelling within the waveguide between two turning points of the helix is $\sqrt{(2\pi r_h)^2 + \Lambda^2}$, and $k_g\sqrt{(2\pi r_h)^2 + \Lambda^2}$ is the corresponding optical phase experienced by the traveling photons.

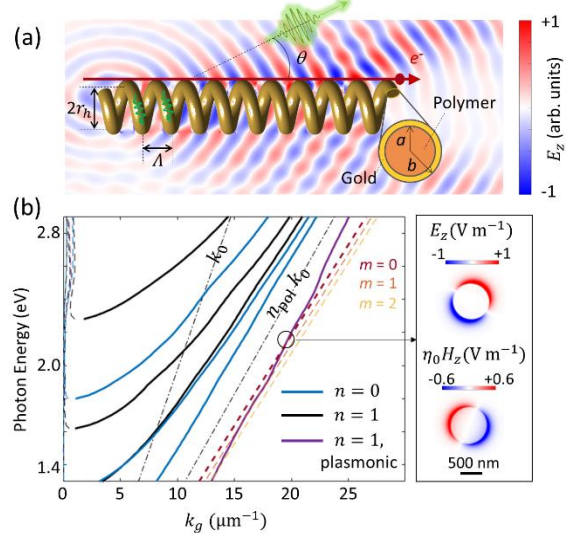

**Figure 1.** Phase-matched generation of Smith-Purcell radiation from a helical waveguide. (a) Topology of the structure, where a moving electron travels parallel with the screw axis of a helix. The helix is composed of a polymer fiber with a radius of $a = 400$ nm and with a 40 nm gold layer deposited on it. (b) Left: Calculated dispersion of the fiber modes. The dashed-dotted lines indicate the light lines in vacuum and fiber. Thin dashed lines show the attenuation constant and thick solid lines the phase constant. The red, orange, and yellow dashed lines specify the phase matching condition corresponding to a moving electron at the kinetic energy of 18 keV (Eq. (1) in the text). Right: Spatial profiles of the z-component of the electric and magnetic fields at the depicted point.

Within the same time interval, the electron travels the distance $\Lambda$ and the incident field corresponding to the near-field of the moving electron undergoes a phase shift equal to $\Lambda\omega/v_e$. In the case of the constructive interaction between the emitted photons at both interaction points, the phase-matching condition leads to $k_g\sqrt{(2\pi r_h)^2 + \Lambda^2} + 2m\pi = \Lambda\omega/v_e$. This condition is expressed as

$$k_g = (\Lambda\omega/v_e - 2m\pi) \cdot \left((2\pi r_h)^2 + \Lambda^2\right)^{-\frac{1}{2}}. \tag{1}$$

The electron-induced guided waves within the waveguide gradually contribute to the radiation due to the sharp bending geometry of the helical structure. Equation (1) is a generalized Smith-Purcell radiation criterion, where the emitted photon paths are modified by accounting for traveling waves inside the waveguide. This generalized phase-matching condition reproduces the Smith-Purcell condition when neglecting the optical path taken by the photons travelling within the waveguide, by inserting $r_h = 0$ and inserting $k_g = k_0 \cos\theta$, where $\theta$ is the angle between the electron trajectory and radiated optical rays. Similarly, it reproduces the Cherenkov condition by inserting $m = 0$ and $r_h = 0$.

The right side of Eq. (1) is provided as red, orange and yellow dashed lines in Fig. 1(b), for an electron at the kinetic energy of 18 keV, and with $m = 0$, 1, and 2, respectively. Electrons having a kinetic energy ($U_{el}$) in the range of $16\,\text{keV} \leq U_{el} \leq 20$ keV are able to couple to the plasmonic mode in a phase-matched condition with $m = 0$, whereas coupling to this mode for higher electron energies is only possible by higher diffraction orders, namely $m > 0$. The metallic layer serves a dual purpose in our configuration: it facilitates the coherent excitation of plasmonic waves and their interaction with electron beams, while also preventing charging of the helix under electron beam irradiation. This merged Cherenkov and Smith-Purcell emission mechanism significantly enhances the radiation. The emitted photons constitute a rate of approximately 0.02 photon per given electron (for a Helix with 11 turns), significantly higher than the emission from planar, plasmonic-based electron-driven photon sources emitting at the rate of approximately $10^{-4}$ per given electron, as confirmed by our measurements presented below. The photon statistics from our EDPHS structures have been obtained by utilizing a photon-counting detector (PCD) connected to our CL setup and calibrating the spectrograph counts by comparing the number of photons obtained via the PCD with the counts registered by the CCD camera, integrated over the spectral range of the EDPHS radiation. The acquisition time was set to either 200 ms or 250 ms, the detector gain was fixed at 1, the binning was 2, and the spectral acquisition was performed with a pixel size ranging from 10 nm to 20 nm.

Notably, in the case of pure Smith–Purcell emission—where the excitation of guided modes is neglected—the emission follows the classical Smith–Purcell condition, as previously discussed. For a 20 keV electron beam to generate light that propagates collinearly with the electron trajectory, the grating pitch must be approximately 224 nm. However, achieving such fine feature sizes exceeds the resolution limits of current 3D fabrication methods based on two-photon polymerization.

The helical waveguides are positioned on top of a plateau fabricated at the same size of the helices (Fig. 2(a) and (b) and supplementary Figure S1). The structures are held by a special holder fabricated for precise alignment with respect to the electron beam, therefore the electron travels parallel with the helix axis (See supplementary Figure S1). For measuring the radiation from electron beams interacting with the helical waveguide, a DELMIC SPARC cathodoluminescence (CL) spectroscopy and angle-resolve mapping system installed in a ZEISS SIGMA scanning electron microscope is used (Fig. S1). A parabolic mirror positioned either above or below the helix gathers the radiation with the collection angular range of $1.46\pi$ sr , and collimates it toward the analyzing path [37]. For the measurements shown below an electron beam at the kinetic energy of 18 keV and the current of 10 nA is used and the mirror is positioned below the helix.

Figure 2(c) shows the CL spectrum as a function of the electron impact position, varying from 320 nm away from the waveguide rim toward the helical axis until a distance of 500 nm from the axis. The intensity reaches a maximum at the photon energy of 2.1 eV at the rim of the waveguide and exhibits an evanescent decay with a decay length of 120 nm (Fig. 2(d)). The emitted photon energy perfectly matches the phase-matching criterion explained by equation (1) and demonstrated in Fig. 1(b). Within the helical arm, the CL intensity also shows an evanescent tail, decreasing to a minimum at a distance of 185 nm from the rim and increasing again towards the opposite rim. At this opposing rim, located at a position 1.2 μm from the origin, the intensity is higher, creating a double-peak feature. This double-peak behavior is attributed to a slight misalignment of the helix, causing it to deviate from being perfectly parallel to the beam trajectory (Fig. S1). The maximum CL intensities observed at the waveguide rims confirm the efficient coupling of the electron beam to the plasmonic mode.

The CL intensity remains at 60% of its maximum amplitude at the helix center, where the electron beam does not interact directly with the material and is approximately 1.3μm away from the waveguide rim. The asymmetrical coupling to the plasmonic modes at the opposing rims, combined with the higher CL intensity observed when the electron travels through the axial void, indicates a significant coupling of the evanescent tails of the plasmonic fields within the axial void. The resulting modal structure facilitates efficient electron-photon interactions, even in regions where the electron is 1μm away from the rim of the waveguide.

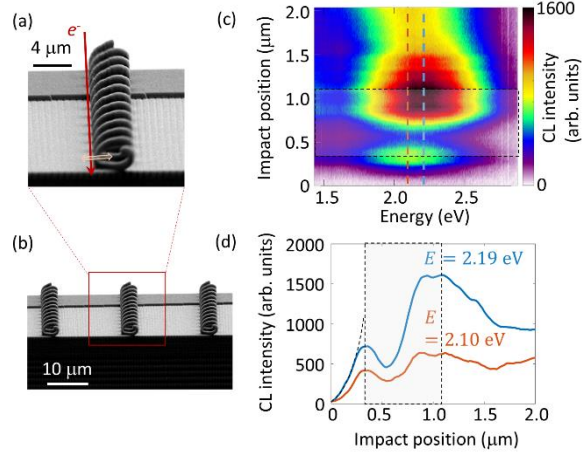

**Figure 2.** 3D-printed helical waveguide. (a) secondary electron image of a single 3D-printed helical waveguide. (b) SE image of a series of such waveguides arranged on a plateau. (c) CL spectra as a function of the electron impact position along the path indicated by the arrow in (a). The region within the core of the helical waveguide and its metallic shell is highlighted with a dashed, shaded box. (d) Linear plots of the CL intensity versus electron impact position at the specified photon energies.

The observed radiation from the helix thus originates from the coupling of the moving electron to a plasmonic mode supported by the gold-coated helical waveguide, rather than to photonic modes confined within the low-index fiber core. In a straight fiber configuration, this coupling leads to guided-mode excitation without far-field emission. However, the helical geometry introduces periodic variations in the interaction distance, enabling phase-matched emission via a combined Smith–Purcell and Cherenkov-like mechanism. The curvature of the helix, with a bending radius approximately three times the effective wavelength of the plasmonic mode, facilitates partial leakage of the guided energy into free space along the waveguide, resulting in directional far-field radiation.

To further illustrate the angular range of the emission, the CL emission is resolved versus the polar angle of the emission and photon energy, by performing momentum-resolved spectroscopy (Fig. 3(a) and (b)) [38]. The azimuthal angle is set to the direction of strongest emission—i.e., $\varphi = \pm 10°$, as confirmed by the angle-resolved emission map presented later in Fig. 3(c).

The angle-resolved spectral CL map displays a broad peak at the energy of 2.1 eV to 2.5 eV, collimated along the symmetry axis of the grating. The dark regions at $\theta = 0$ and at $\theta > 80°$ correspond to the hole implemented into the mirror for avoiding electron beam contamination and damage, and the restricted collection efficiency of the mirror due to the its limited size, respectively. This collimated emission and its photon energy agrees well with the $m = 0$ emission (dashed red line in Fig. 1(b)). The higher-order diffraction orders are obvious as well with faint directional emissions at higher angles, marked by dashed orange ($m = 1$) and dashed yellow ($m = 2$) lines in Fig. 3(b).

The excitation of the helix with the electron beam parallel to the helix axis leads to a chiral polarization following the helicity of the structure. The emission profile therefore follows as well the handedness of the helix, in such a way that the emission is predominantly in the form of right-handed circularly polarized light, where the detector convention is considered. Measured stokes parameters, namely $S_0 = |A_x|^2 + |A_y|^2$ (light intensity), $S_1 = |A_x|^2 - |A_y|^2$, and $S_3 = 2\operatorname{Im}\{A_x^* A_y\}$ verify the generation of directional and right-handed circularly-polarized light (Fig. 3(c)). In the above equations, $A_x$ and $A_y$ depict the complex amplitudes of the $x$- and $y$-components of the electric field. For an elliptically-polarized light, where a phase offset exists between $A_x$ and $A_y$ components, $S_3$ takes a non-zero value and its sign is associated with the handedness of the polarization (positive for right-handed and negative for left-handed polarization, when the detector convention is used).

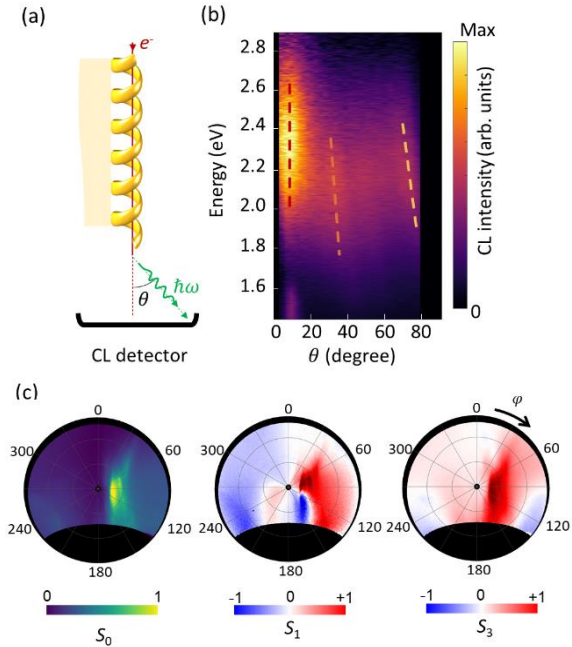

**Figure 3.** (a) The schematic of the experiment. (b) Angle-resolved CL spectra, and (c) angle-resolved stokes parameters of the radiation measured at the integrated photon energy range of 2.15 eV to 2.26 eV. Electron beam has the kinetic energy of 18 keV and propagates parallel to the helix axis at the impact position of $r_h + b + 20\,\mathrm{nm}$, forming a spot size of 100 nm.

To further demonstrate that the radiation originates from the combined Cherenkov and Smith-Purcell emission mechanism described by Eq. (1), we have performed angle-resolved spectroscopy and polarimetry measurements for the case where the electron beam traverses the helix perpendicular to its axis (Fig. 4). In this configuration, the emission exhibits a distinct profile, peaking at 1.6 eV and extending toward higher angular ranges. The emitted photon energy in this case does not depend on the electron impact position. Since the helix is long, reflections from its ends are negligible, preventing the formation of standing-wave patterns within the helix. This indicates that the electron primarily couples to lower-energy modes of the waveguide, which emit from the waveguide's ends, resulting in a broad angular distribution.

Polarimetry measurements further reveal significant differences in polarization between emissions from the two ends of the waveguide. The left end predominantly emits right-handed polarized light, while the right end emits left-handed polarized light. These results align with observations reported elsewhere and can be attributed to the dependence of the emission's chirality on the electron beam's impact position [39].

The emergence of circular polarization in the emitted light can be attributed primarily to the geometrical chirality of the helical waveguide. While the supported plasmonic mode possesses an azimuthal dependence (mode number $n$ = 1; see Fig.1b), which reflects its orbital angular momentum, its polarization state is not directly determined by this modal property of the straight fiber. Instead, the helical geometry induces a rotating plasmonic surface current, which introduces a dynamic phase shift between the transverse components of the radiated field. This mechanism acts analogously to a quarter-wave plate, effectively generating circular polarization. Experimental observations and numerical simulations confirm that the handedness of the helix governs the handedness of the emitted circularly polarized light.

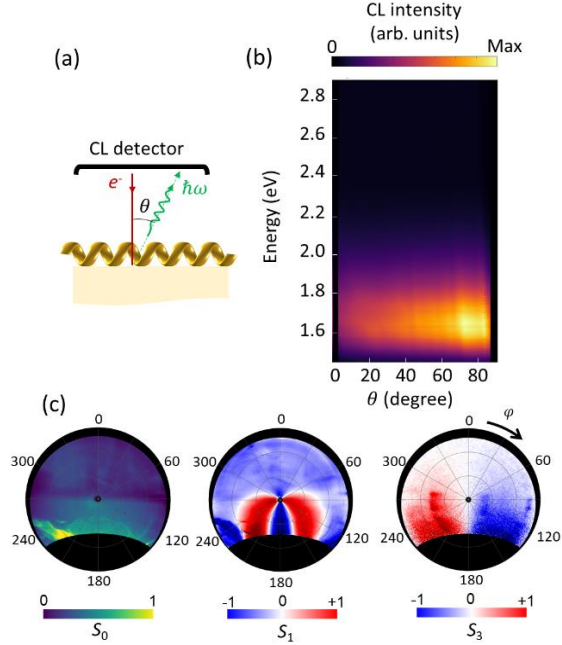

**Figure 4.** (a) Schematic of the experiment. (b) Angle-resolved CL spectra and (c) angle-resolved stokes parameters of the radiation measured at the integrated photon energy range of 1.6 eV to 1.7 eV. The electron beam has the kinetic energy of 18 keV and propagates perpendicular to the helix axis.

Noticeably, the emission from the helix in the case when electron travels parallel to the helix axis, cannot be described by pure Smith-Purcell radiation. There is a large mismatch between the momentum of the free-space light and the momentum of the current-density distribution of the moving electron at the kinetic energy of 18 keV. Therefore, to satisfy the Smith-Purcell criterion, namely, $\omega v_e^{-1} = k_0 \cos\theta + 2m\pi\Lambda^{-1}$, diffraction orders of $13 \leq m \leq 35$ is required, for the photon energy within 1 eV and 3 eV, which makes the interaction rather inefficient.

The simulations of a moving electron interacting with the helix reproduce the main observed features. These simulations were performed using a finite-difference time-domain (FDTD) solver developed in our group [40]. The large size of the structure, combined with its fine features (e.g., the gold layer thickness and confined plasmon polaritons), posed significant computational challenges. We discretised the entire volume using a 15 nm grid size and introduced 20000 time iterations. A second-order radiation boundary condition was employed to eliminate the need for additional volume required by perfectly matched layers. The calculated spectrum of the light intensity (Poynting vector) at surface positioned 8 μm below the helix demonstrates a broad resonance centered at $E = 2.2$ eV in a good agreement with observed experimental results. The emission from the helix is also collimated downward, making an angle of approximately 20 degrees with the electron beam trajectory. The transverse field intensity ($|E_x|^2 + |E_y|^2$) has a collimated spatial profile and constitutes an elliptical polarization (Inset of Fig. 5(a)).

The field profiles are symmetrical distributed with respect to the helix axis, in contrast with the experiments, due to the presence of the plateau in the latter that is required to hold the helix.

Similar to the Cherenkov radiation and the Smith-Purcell effect, the energy of the emitted photons can be tuned via the electron kinetic energy. Supplementary Note S3 and Fig. S2 indicate that this indeed can be observed as a control parameter.

Based on comparisons between the experimental results and the phase-matching condition—where the phase constant of the optical modes is assumed to correspond to that of a straight optical fiber—we conclude that the optical modes are only slightly modified by transforming the waveguide into a helical structure. This is justified by the relatively large helical pitch of the helix compared to the fiber's radius, which supports this assumption.

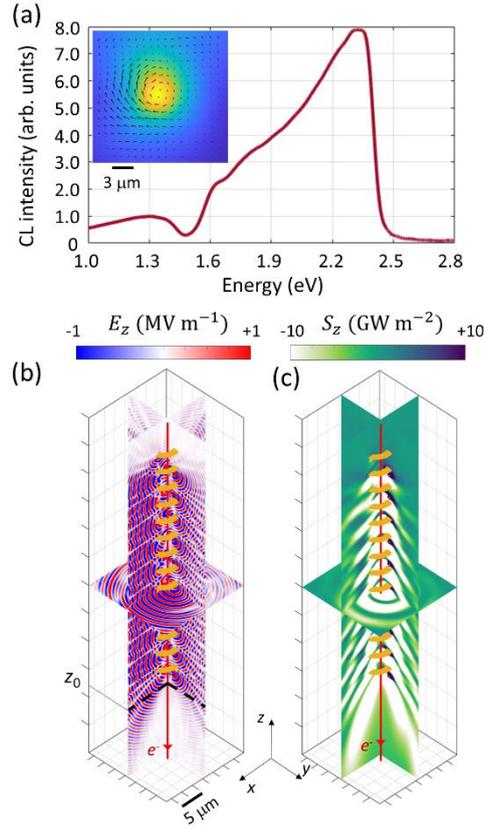

**Figure 5.** Simulation results, exhibiting the interaction of a moving electron at the kinetic energy of 18 eV interacting with the helix. The inset shows the spatial profile of the *z*-component of the electric field at $z = z_0$. The inset displays the spatial profile of absolute value of the transverse electric field at the xy-plane marked by $z_0$, positioned at $3\,\mu m$ below the helix end. (a) Radiation spectrum, calculated at the *xy*-plane positioned at $8\,\mu m$ below the helix. The spatial profile of the (b) *z*-component of the electric field and (c) the z-component of the Poynting vector, at $E = 2.22\,\mu m$.

The performance of the helical waveguides—particularly emission efficiency and polarization characteristics—can be influenced by fabrication imperfections such as surface roughness or slight asymmetries. To evaluate the impact of these tolerances, we conducted cathodoluminescence measurements across multiple fabricated helices. The consistency observed in spectral profiles, angular distributions, and polarization states suggests that the emission characteristics are robust with respect to minor fabrication variations.

Finally, helical structures have previously been proposed to generate helically shaped optical wavefronts. However, our helical waveguide takes advantage of the guided modes of a 3D-printed waveguide to generate phase-matched radiation from electron beams in the visible range, unlike the generation of terahertz Smith-Purcell radiation [41-43].

**Conclusion**

In conclusion, we have established a new phase-matching scenario that enables stronger electron-photon interactions. This is achieved by maintaining synchronicity between the waveguiding modes of an optical fiber made into a helix and electron beams through the introduction of extended optical paths and multiple sequential interactions. The structure is fabricated using two-photon-polymerization-based 3D printers (Quantum X series,

NanoScribe Company), enabling a new paradigm for strong electron-photon interactions. Our phase-matching scenario could be modified for even stronger interactions, e. g., by inserting the helix in a hollow cylindrical waveguide. In an inverse approach [44-46], launching optical waves into the fiber and leveraging laser-induced near-field interactions could enable regimes of ultra-strong electron-photon interactions, which are highly favorable for dielectric laser accelerators.

**Acknowledgments**

We acknowledge collaborations with Dr. Jochen Zimmer, Mr. Fabian Rahn and Dr. Aaron Kobler for fabricating the structures (NanoScribe Company). This project has received funding from the Volkswagen Foundation (Momentum Grant), European Research Council (ERC) under the European Union's Horizon 2020 research and innovation program under grant agreement no. 101017720 (EBEAM), from the European Research Council (ERC Consolidator Grant UltraSpecT with no. 101170341; ERC Proof-of-Concept Grant UltraCoherentCL with no. 101157312) and from Deutsche Forschungsgemeinschaft.

**References**

[1] N. Talebi *et al.*, "Merging transformation optics with electron-driven photon sources," *Nature Communications,* vol. 10, no. 1, p. 599, 2019/02/05 2019, doi: 10.1038/s41467-019-08488-4.
[2] J. Christopher, M. Taleb, A. Maity, M. Hentschel, H. Giessen, and N. Talebi, "Electron-driven photon sources for correlative electron-photon spectroscopy with electron microscopes," *Nanophotonics,* vol. 9, no. 15, pp. 4381-4406, 2020, doi: doi:10.1515/nanoph-2020-0263.
[3] N. Talebi, "Interaction of electron beams with optical nanostructures and metamaterials: from coherent photon sources towards shaping the wave function," *Journal of Optics,* vol. 19, no. 10, p. 103001, 2017/09/12 2017, doi: 10.1088/2040-8986/aa8041.
[4] G. Adamo *et al.*, "Light Well: A Tunable Free-Electron Light Source on a Chip," *Physical Review Letters,* vol. 103, no. 11, p. 113901, 09/09/ 2009, doi: 10.1103/PhysRevLett.103.113901.
[5] N. Talebi, "Spectral Interferometry with Electron Microscopes," *Scientific Reports,* vol. 6, no. 1, p. 33874, 2016/09/21 2016, doi: 10.1038/srep33874.
[6] M. Taleb, M. Hentschel, K. Rossnagel, H. Giessen, and N. Talebi, "Phase-locked photon–electron interaction without a laser," *Nature Physics,* vol. 19, no. 6, pp. 869-876, 2023/06/01 2023, doi: 10.1038/s41567-023-01954-3.
[7] P. B. Masoud Taleb, Mximilian Black, Mario Hentschel, Wilfried Sigle, Benedikt Haas, Christoph Koch, Peter A. van Aken, Harald Giessen, Nahid Talebi, "Ultrafast phonon-mediated dephasing of color centers in hexagonal boron nitride probed by electron beams," *arXiv:2404.09879* 2024.
[8] M. Taleb *et al.*, "Ultrafast phonon-mediated dephasing of color centers in hexagonal boron nitride probed by electron beams," *Nature Communications,* vol. 16, no. 1, p. 2326, 2025/03/08 2025, doi: 10.1038/s41467-025-57584-1.
[9] N. van Nielen, M. Hentschel, N. Schilder, H. Giessen, A. Polman, and N. Talebi, "Electrons Generate Self-Complementary Broadband Vortex Light Beams Using Chiral Photon Sieves," *Nano Letters,* vol. 20, no. 8, pp. 5975-5981, 2020/08/12 2020, doi: 10.1021/acs.nanolett.0c01964.
[10] G. Li, B. P. Clarke, J.-K. So, K. F. MacDonald, and N. I. Zheludev, "Holographic free-electron light source," *Nature Communications,* vol. 7, no. 1, p. 13705, 2016/12/02 2016, doi: 10.1038/ncomms13705.
[11] N. Talebi *et al.*, "Excitation of Mesoscopic Plasmonic Tapers by Relativistic Electrons: Phase Matching versus Eigenmode Resonances," *ACS Nano,* vol. 9, no. 7, pp. 7641-7648, 2015/07/28 2015, doi: 10.1021/acsnano.5b03024.
[12] J. McNeur *et al.*, "Elements of a dielectric laser accelerator," *Optica,* vol. 5, no. 6, pp. 687-690, 2018/06/20 2018, doi: 10.1364/OPTICA.5.000687.
[13] L. Zhang, H. Xu, and W. Liu, "Inverse Cherenkov dielectric laser accelerator with alternating phase focusing for subrelativistic particles," *Physical Review Accelerators and Beams,* vol. 27, no. 11, p. 110401, 11/13/ 2024, doi: 10.1103/PhysRevAccelBeams.27.110401.
[14] A. Feist *et al.*, "Cavity-mediated electron-photon pairs," *Science,* vol. 377, no. 6607, pp. 777-780, 2022, doi: doi:10.1126/science.abo5037.
[15] R. Dahan *et al.*, "Resonant phase-matching between a light wave and a free-electron wavefunction," *Nature Physics,* vol. 16, no. 11, pp. 1123-1131, 2020/11/01 2020, doi: 10.1038/s41567-020-01042-w.

[16] O. Kfir et al., "Controlling free electrons with optical whispering-gallery modes," *Nature,* vol. 582, no. 7810, pp. 46-49, 2020/06/01 2020, doi: 10.1038/s41586-020-2320-y.
[17] I. M. Frank, "A conceptual history of the Vavilov-Cherenkov radiation," *Soviet Physics Uspekhi,* vol. 27, no. 5, p. 385, 1984/05/31 1984, doi: 10.1070/PU1984v027n05ABEH004300.
[18] S. J. Smith and E. M. Purcell, "Visible Light from Localized Surface Charges Moving across a Grating," *Physical Review,* vol. 92, no. 4, pp. 1069-1069, 11/15/ 1953, doi: 10.1103/PhysRev.92.1069.
[19] F. J. García de Abajo, "Optical excitations in electron microscopy," *Reviews of Modern Physics,* vol. 82, no. 1, pp. 209-275, 02/03/ 2010, doi: 10.1103/RevModPhys.82.209.
[20] Y. Adiv et al., "Observation of 2D Cherenkov Radiation," *Physical Review X,* vol. 13, no. 1, p. 011002, 01/06/ 2023, doi: 10.1103/PhysRevX.13.011002.
[21] F. J. García de Abajo, A. Rivacoba, N. Zabala, and N. Yamamoto, "Boundary effects in Cherenkov radiation," *Physical Review B,* vol. 69, no. 15, p. 155420, 04/19/ 2004, doi: 10.1103/PhysRevB.69.155420.
[22] Z. C. Zetao Xie, Hao Li, Qinghui Yan, Hongsheng Chen, Xiao Lin, Ido Kaminer, Owen D. Miller, Yi Yang, "Maximal quantum interaction between free electrons and photons," *arXiv:2404.00377* 2024.
[23] N. Talebi, "A directional, ultrafast and integrated few-photon source utilizing the interaction of electron beams and plasmonic nanoantennas," *New Journal of Physics,* vol. 16, no. 5, p. 053021, 2014/05/09 2014, doi: 10.1088/1367-2630/16/5/053021.
[24] Y. J. Tan, P. Pitchappa, N. Wang, R. Singh, and L. J. Wong, "Space-Time Wave Packets from Smith-Purcell Radiation," *Advanced Science,* vol. 8, no. 22, p. 2100925, 2021, doi: https://doi.org/10.1002/advs.202100925.
[25] P. Zhang, Y. Zhang, and M. Tang, "Enhanced THz Smith-Purcell radiation based on the grating grooves with holes array," *Opt. Express,* vol. 25, no. 10, pp. 10901-10910, 2017/05/15 2017, doi: 10.1364/OE.25.010901.
[26] J. H. Brownell, J. Walsh, and G. Doucas, "Spontaneous Smith-Purcell radiation described through induced surface currents," *Physical Review E,* vol. 57, no. 1, pp. 1075-1080, 01/01/ 1998, doi: 10.1103/PhysRevE.57.1075.
[27] R. Remez et al., "Spectral and spatial shaping of Smith-Purcell radiation," *Physical Review A,* vol. 96, no. 6, p. 061801, 12/06/ 2017, doi: 10.1103/PhysRevA.96.061801.
[28] Z. Rezaei and B. Farokhi, "Start current and growth rate in Smith–Purcell free-electron laser with dielectric-loaded cylindrical grating," *Journal of Theoretical and Applied Physics,* vol. 14, no. 2, pp. 149-158, 2020/06/01 2020, doi: 10.1007/s40094-019-00358-0.
[29] R. Rong et al., "The Interaction of 2D Materials With Circularly Polarized Light," *Advanced Science,* vol. 10, no. 10, p. 2206191, 2023, doi: https://doi.org/10.1002/advs.202206191.
[30] J. Giltinan, V. Sridhar, U. Bozuyuk, D. Sheehan, and M. Sitti, "3D Microprinting of Iron Platinum Nanoparticle-Based Magnetic Mobile Microrobots," *Advanced Intelligent Systems,* vol. 3, no. 1, p. 2000204, 2021, doi: https://doi.org/10.1002/aisy.202000204.
[31] B. Jian, H. Li, X. He, R. Wang, H. Y. Yang, and Q. Ge, "Two-photon polymerization-based 4D printing and its applications," *International Journal of Extreme Manufacturing,* vol. 6, no. 1, p. 012001, 2023/10/06 2024, doi: 10.1088/2631-7990/acfc03.
[32] F. Rothermel et al., "Fabrication and Characterization of a Magnetic 3D-printed Microactuator," *Advanced Materials Technologies,* vol. 9, no. 12, p. 2302196, 2024, doi: https://doi.org/10.1002/admt.202302196.
[33] L. Siegle, S. Ristok, and H. Giessen, "Complex aspherical singlet and doublet microoptics by grayscale 3D printing," *Opt. Express,* vol. 31, no. 3, pp. 4179-4189, 2023/01/30 2023, doi: 10.1364/OE.480472.
[34] F. Mayer et al., "3D Two-Photon Microprinting of Nanoporous Architectures," *Advanced Materials,* vol. 32, no. 32, p. 2002044, 2020, doi: https://doi.org/10.1002/adma.202002044.
[35] V. Hahn et al., "Light-sheet 3D microprinting via two-colour two-step absorption," *Nature Photonics,* vol. 16, no. 11, pp. 784-791, 2022/11/01 2022, doi: 10.1038/s41566-022-01081-0.
[36] *See Supplemental Material [url], which includes Refs. [42-44].*
[37] T. Coenen, B. J. M. Brenny, E. J. Vesseur, and A. Polman, "Cathodoluminescence microscopy: Optical imaging and spectroscopy with deep-subwavelength resolution," *MRS Bulletin,* vol. 40, no. 4, pp. 359-365, 2015, doi: 10.1557/mrs.2015.64.
[38] S. Mignuzzi et al., "Energy–Momentum Cathodoluminescence Spectroscopy of Dielectric Nanostructures," *ACS Photonics,* vol. 5, no. 4, pp. 1381-1387, 2018/04/18 2018, doi: 10.1021/acsphotonics.7b01404.
[39] R. Lingstädt et al., "Electron Beam Induced Circularly Polarized Light Emission of Chiral Gold Nanohelices," *ACS Nano,* vol. 17, no. 24, pp. 25496-25506, 2023/12/26 2023, doi: 10.1021/acsnano.3c09336.
[40] N. Talebi, W. Sigle, R. Vogelgesang, and P. van Aken, "Numerical simulations of interference effects in photon-assisted electron energy-loss spectroscopy," *New Journal of Physics,* vol. 15, no. 5, p. 053013, 2013/05/08 2013, doi: 10.1088/1367-2630/15/5/053013.

[41] J.-F. Zhu, C.-H. Du, Z.-W. Zhang, P.-K. Liu, L. Zhang, and A. W. Cross, "Smith–Purcell radiation from helical grating to generate wideband vortex beams," *Opt. Lett.,* vol. 46, no. 18, pp. 4682-4685, 2021/09/15 2021, doi: 10.1364/OL.434794.

[42] L. Jing *et al.*, "Spiral Field Generation in Smith-Purcell Radiation by Helical Metagratings," *Research,* vol. 2019, 2019, doi: doi:10.34133/2019/3806132.

[43] A. R. Neureuther and R. Mittra, "Smith-Purcell radiation from a narrow tape helix," *Proceedings of the IEEE,* vol. 55, no. 12, pp. 2134-2142, 1967, doi: 10.1109/PROC.1967.6091.

[44] K. Mizuno, J. Pae, T. Nozokido, and K. Furuya, "Experimental evidence of the inverse Smith–Purcell effect," *Nature,* vol. 328, no. 6125, pp. 45-47, 1987/07/01 1987, doi: 10.1038/328045a0.

[45] N. Talebi, "Schrödinger electrons interacting with optical gratings: quantum mechanical study of the inverse Smith–Purcell effect," *New Journal of Physics,* vol. 18, no. 12, p. 123006, 2016/12/02 2016, doi: 10.1088/1367-2630/18/12/123006.

[46] R. Shiloh *et al.*, "Electron phase-space control in photonic chip-based particle acceleration," *Nature,* vol. 597, no. 7877, pp. 498-502, 2021/09/01 2021, doi: 10.1038/s41586-021-03812-9.

Supplementary Information for:

# Phase-matched electron-photon interactions enabled with 3D-printed helical waveguides

Masoud Taleb,[1] Mohsen Samadi,[2] and Nahid Talebi[1,*]

[1]Institute of Experimental and Applied Physics, Kiel University, 24098 Kiel, Germany
[2]Department of Electrical and Information Engineering, Kiel University, 24143Kiel, Germany

Content:
1- Optical modes of the polymer/gold fiber
2- Images of the helical waveguide and sample holder
3- Dependence of the emission wavelength on the electron's kinetic energy

### 1- Optical modes of the polymer/gold fiber

To construct the optical modes of the system, we employ the vector potential approach in the cylindrical coordinate system, with $\rho$, $\varphi$, and $z$ being the radius in *xy*-plane, azimuthal angle, and the *z*-axis respectively [1]. Optical modes in fibers exhibit a hybrid nature; that is, the modes are not purely transverse except when $n = 0$, where *n* represents the azimuthal degree of freedom [2]. The fiber is composed of a polymer core with the permittivity $\varepsilon_{r1} = 2.63$ within the region specified by $\rho < a$ and a gold thin cladding with the permittivity $\varepsilon_{r2}$ [3] within the region $a < \rho < b$, whereas the region $\rho > b$ is considered to be vacuum with $\varepsilon_{r3} = 1$. The solution Ansatz constitutes spatial distributions of the magnetic vector potential, as

$$A_z = \begin{cases} C_1 I_n(\kappa_1 \rho) e^{in\varphi} e^{-ik_z z} & \rho < a \\ (C_2 I_n(\kappa_2 \rho) + C_3 K_n(\kappa_2 \rho)) e^{in\varphi} e^{-ik_z z} & a < \rho < b, \\ C_4 K_n(\kappa_3 \rho) e^{in\varphi} e^{-ik_z z} & \rho > b \end{cases} \quad (S1)$$

and the electric vector potential, as

$$F_z = \begin{cases} D_1 I_n(\kappa_1 \rho) e^{in\varphi} e^{-ik_z z} & \rho < a \\ (D_2 I_n(\kappa_2 \rho) + D_3 K_n(\kappa_2 \rho)) e^{in\varphi} e^{-ik_z z} & a < \rho < b, \\ D_4 K_n(\kappa_3 \rho) e^{in\varphi} e^{-ik_z z} & \rho > b \end{cases} \quad (S2)$$

where $C_i$ and $D_i$ are unknown coefficients to be obtained via satisfying the boundary conditions. $k_z$ is the propagation constant of the waves propagating along the $z$ – axis, and the characteristic equations in all three regions are $-\kappa_i^2 + k_z^2 = \varepsilon_{ri} k_0^2$, with $i$ = 1, 2, and 3. $I_n$ and $K_n$ are the modified Bessel functions of the first and second kinds with order $n$.

The electric and magnetic field coefficients are obtained as

$$\vec{E}(\vec{r}) = -\vec{\nabla} \times \vec{F} + (i\omega\varepsilon_0 \varepsilon_r)^{-1} (\vec{\nabla} \times \vec{\nabla} \times \vec{A}) \tag{S3}$$

and

$$\vec{H}(\vec{r}) = +\vec{\nabla} \times \vec{A} + (i\omega\mu_0)^{-1} (\vec{\nabla} \times \vec{\nabla} \times \vec{F}), \tag{S4}$$

Respectively, with both $\vec{A}$ and $\vec{F}$ vectors composed of only z-components. After obtaining the field coefficients by using equations (S1) to (S2), the tangential boundary conditions are satisfied, and therefore the following 4 coupled equations are obtained that relates $C_2$, $C_3$, $D_2$, and $D_3$ as

$$\begin{aligned}
&+\left[\kappa_2^2 - \kappa_1^2\right] \frac{1}{i\omega\varepsilon_0 \varepsilon_{r2}} \frac{nk_z}{a} I_n(\kappa_2 a) I_n(\kappa_1 a) C_2 + \left[\kappa_2^2 - \kappa_1^2\right] \frac{1}{i\omega\varepsilon_0 \varepsilon_{r2}} \frac{nk_z}{a} K_n(\kappa_2 a) I_n(\kappa_1 a) C_3 \\
&+\kappa_1 \kappa_2 \left\{\kappa_2 I_n'(\kappa_1 a) I_n(\kappa_2 a) - \kappa_1 I_n'(\kappa_2 a) I_n(\kappa_1 a)\right\} D_2 \\
&+\kappa_1 \kappa_2 \left\{\kappa_2 I_n'(\kappa_1 a) K_n(\kappa_2 a) - \kappa_1 K_n'(\kappa_2 a) I_n(\kappa_1 a)\right\} D_3 = 0,
\end{aligned} \tag{S5}$$

$$\begin{aligned}
&+\left[\kappa_2^2 - \kappa_3^2\right] \frac{1}{i\omega\varepsilon_0 \varepsilon_{r2}} \frac{nk_z}{b} I_n(\kappa_2 b) K_n(\kappa_3 b) C_2 + \left[\kappa_2^2 - \kappa_3^2\right] \frac{1}{i\omega\varepsilon_0 \varepsilon_{r2}} \frac{nk_z}{b} K_n(\kappa_2 b) K_n(\kappa_3 b) C_3 \\
&+\kappa_2 \kappa_3 \left[\kappa_2 K_n'(\kappa_3 b) I_n(\kappa_2 b) - \kappa_3 I_n'(\kappa_2 b) K_n(\kappa_3 b)\right] D_2 \\
&+\kappa_2 \kappa_3 \left[\kappa_2 K_n'(\kappa_3 b) K_n(\kappa_2 b) - \kappa_3 K_n'(\kappa_2 b) K_n(\kappa_3 b)\right] D_3 = 0,
\end{aligned} \tag{S6}$$

$$\begin{aligned}
&\kappa_1 \kappa_2 \left[-\frac{\varepsilon_{r1}}{\varepsilon_{r2}} \kappa_2 I_n'(\kappa_1 a) I_n(\kappa_2 a) + \kappa_1 I_n'(\kappa_2 a) I_n(\kappa_1 a)\right] C_2 \\
&+\kappa_1 \kappa_2 \left[-\frac{\varepsilon_{r1}}{\varepsilon_{r2}} \kappa_2 I_n'(\kappa_1 a) K_n(\kappa_2 a) + \kappa_1 K_n'(\kappa_2 a) I_n(\kappa_1 a)\right] C_3 \\
&+\left[\kappa_2^2 - \kappa_1^2\right] \frac{1}{i\omega\mu_0} \frac{nk_z}{a} I_n(\kappa_2 a) I_n(\kappa_1 a) D_2 + \left[\kappa_2^2 - \kappa_1^2\right] \frac{1}{i\omega\mu_0} \frac{nk_z}{a} K_n(\kappa_2 a) I_n(\kappa_1 a) D_3 = 0,
\end{aligned} \tag{S7}$$

and

$$\begin{aligned}
&\kappa_2 \kappa_3 \left[-\frac{\varepsilon_{r3}}{\varepsilon_{r2}} \kappa_2 K_n'(\kappa_3 b) I_n(\kappa_2 b) + \kappa_3 I_n'(\kappa_2 b) K_n(\kappa_3 b)\right] C_2 \\
&+\kappa_2 \kappa_3 \left[-\frac{\varepsilon_{r3}}{\varepsilon_{r2}} \kappa_2 K_n'(\kappa_3 b) K_n(\kappa_2 b) + \kappa_3 K_n'(\kappa_2 b) K_n(\kappa_3 b)\right] C_3 \\
&+\left[\kappa_2^2 - \kappa_3^2\right] \frac{1}{i\omega\mu_0} \frac{nk_z}{b} I_n(\kappa_2 b) K_n(\kappa_3 b) D_2 + \left[\kappa_2^2 - \kappa_3^2\right] \frac{1}{i\omega\mu_0} \frac{nk_z}{b} K_n(\kappa_2 b) K_n(\kappa_3 b) D_3 = 0.
\end{aligned} \tag{S8}$$

Equations (S5) to (S8) defines a nonlinear Eigenvalue problem, for obtaining the propagation constant $k_z$ and the eigenvectors $C_2$, $C_3$, $D_2$, and $D_3$. The unknown coefficients $C_1$, $C_4$, $D_1$, and $D_4$ are obtained as

$$C_1 = +\frac{\varepsilon_{r1}}{\varepsilon_{r2}} \frac{\kappa_2^2}{\kappa_1^2} \frac{I_n(\kappa_2 a)}{I_n(\kappa_1 a)} C_2 + \frac{\varepsilon_{r1}}{\varepsilon_{r2}} \frac{\kappa_2^2}{\kappa_1^2} \frac{K_n(\kappa_2 a)}{I_n(\kappa_1 a)} C_3, \tag{S9}$$

$$C_4 = +\frac{\varepsilon_{r3}}{\varepsilon_{r2}} \frac{\kappa_2^2}{\kappa_3^2} \frac{I_n(\kappa_2 b)}{K_n(\kappa_3 b)} C_2 + \frac{\varepsilon_{r3}}{\varepsilon_{r2}} \frac{\kappa_2^2}{\kappa_3^2} \frac{K_n(\kappa_2 b)}{K_n(\kappa_3 b)} C_3, \tag{S10}$$

$$D_1 = +D_2 \frac{\kappa_2^2}{\kappa_1^2} \frac{I_n(\kappa_2 a)}{I_n(\kappa_1 a)} + D_3 \frac{\kappa_2^2}{\kappa_1^2} \frac{K_n(\kappa_2 a)}{I_n(\kappa_1 a)}, \tag{S11}$$

and

$$D_4 = +D_2 \frac{\kappa_2^2}{\kappa_3^2} \frac{I_n(\kappa_2 b)}{K_n(\kappa_3 b)} + D_3 \frac{\kappa_2^2}{\kappa_3^2} \frac{K_n(\kappa_2 b)}{K_n(\kappa_3 b)},$$

Which are used to calculate the spatial distribution of the electric and magnetic fields in all regions.

## 2. Images of the helical waveguide and sample holder

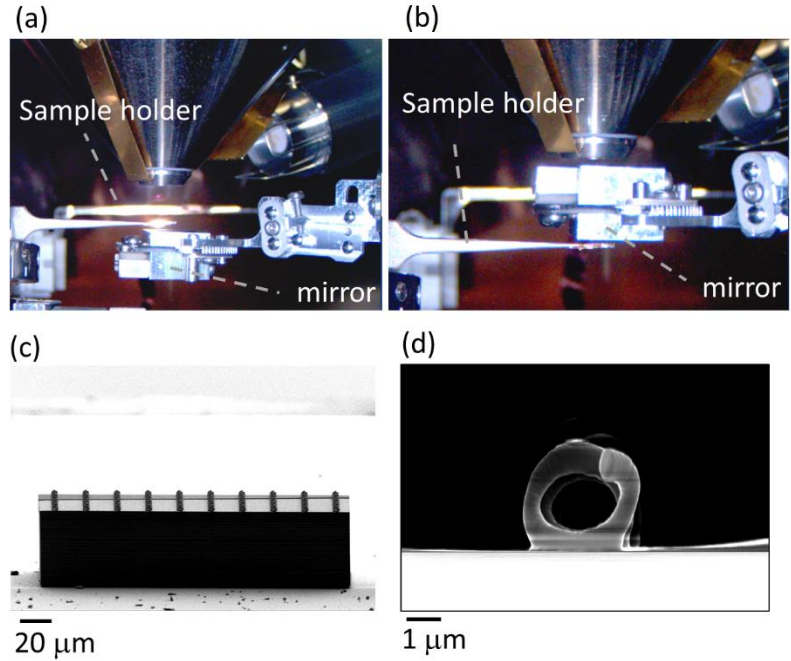

**FIG. S1.** (a) The setup, consisting of a sample holder inside a scanning electron microscope and a parabolic mirror positioned below the sample. (b) The same setup with the parabolic mirror positioned above the sample. (c) A scanning electron (SE) image of the sample, showing a series of microhelices arranged on top of a plateau. (d) A high-magnification top-view image of a single helix, demonstrating the alignment of the helix axis parallel to the electron trajectory.

## 3. Dependence of the emission wavelength on the electron's kinetic energy

The phase-matching condition defined in Eq. (1) of the main text enables control over the photon energy and intensity for a specific helix through various parameters, including the electron's group velocity and the diffraction order $m$. Fig. S2 illustrates that the emitted photon energy is indeed dependent on the electron's kinetic energy. For an electron with the kinetic energy of 15 keV ($v_e = 0.24c$, where $c$ is the light speed in vacuum), only a faint emission is observed, which is two orders of magnitude weaker than the emission from a 17 keV electron beam. For the latter, the emission occurs at the peak photon energy of $E = 2.2$ eV, in a good agreement with the phase-matching condition. For an electron beam with the kinetic energy of 20 keV ($v_e = 0.27c$), the peak photon energy occurs at $E = 2.4$ eV, whereas the phase-matching condition specifies an emission at the energy of 2.65 eV. The emission angle is higher than that observed for an electron with a kinetic energy of 17 keV. Both observations indicate that the emission corresponds to the $m = 1$ diffraction order. This condition results in an emission at a photon energy of 2.5 eV, which aligns better with the experimental observations.

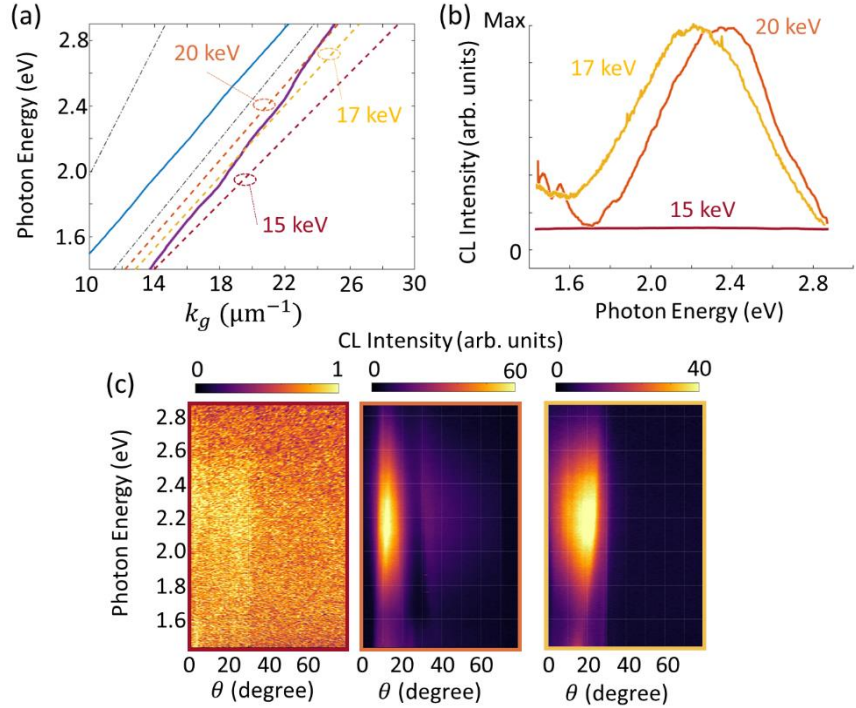

**FIG. 3.** (a) Dispersion diagram of the fundamental mode (Solid purple line) and the second mode (solid blue line) of the optical fiber. Dashed-dotted lines display the optical lines in vacuum and in the polymer. The colored dashed lines exhibit the phase-matching condition for an electron with depicted kinetic energies propagating parallel to the helix axis. (b) CL spectra and (b) CL angle-resolved spectral maps for an electron with the kinetic energies of 15 keV (left), 17 keV (middle), and 20 keV (right) propagating parallel to the helix.

**References:**
[1] R. F. Harrington, *Time-harmonic electromagnetic fields* (mcGraw-Hill Book Company, New York, 1961).
[2] A. H. Cherin, *Introduction to Optical Fibers* (McGraw-Hill, US, 1982).
[3] P. B. Johnson and R. W. Christy, Physical Review B **6**, 4370 (1972).